# Multiagent Reinforcement Learning Enhanced Decision-making of Crew Agents During Floor Construction Process

Bin Yang [a], Boda Liu [a,*], Yilong Han [b,**], Xin Meng [a], Yifan Wang[a], Hansi Yang [a], Jianzhuang Xiao [a,c]

[1] *College of Civil Engineering, Tongji University, Shanghai, China*

[2] *School of Economics & Management, Tongji University, Shanghai, China*

[3] *Guangxi University, Nanning, China*

* Corresponding author, E-mail address :w-lbd-w@tongji.edu.cn

* * Corresponding author, E-mail address : yilong.han@tongji.edu.cn

**Abstract**

Fine-grained simulation of floor construction processes is essential for supporting lean management and the integration of information technology. However, existing research does not adequately address the on-site decision-making of constructors in selecting tasks and determining their sequence within the entire construction process. Moreover, decision-making frameworks from computer science and robotics are not directly applicable to construction scenarios. To facilitate intelligent simulation in construction, this study introduces the Construction Markov Decision Process (CMDP). This approach transforms construction activities, which involve multiple agents with subjective decisions, into a structured process of target evaluation, selection, and execution. By integrating construction knowledge, the CMDP enables training of an agent's hierarchical policy to efficiently identify and achieve high-value construction objectives, aligning with project requirements. The CMDP is developed on the Unity platform, utilizing a two-stage training approach with the Multi-agent proximal policy optimization algorithm. A case study demonstrates the effectiveness of this framework: the low-level policy successfully simulates the construction process, while the high-level policy develops a strategy that alternates between dense and spacious areas, thus significantly improving construction efficiency in various dimensions.

**Keywords:** Multi-agent reinforcement learning, Agent-based modeling, Construction management, Construction Markov Decision Process (CMDP)

# 1. Introduction

The construction process is a multifaceted system characterized by a range of procedures and intricate dynamic interactions. The rapid advancement of information and artificial intelligence technologies has enabled researchers to detect and understand construction activities at the level of worker actions [1], access their interactions with equipment [2], and simulate pre-planned tasks for robotic implementation [3]. However, the real-time decision-making simulation and optimization involving construction performers choosing among different tasks remains underexplored. The movements of construction performers not only initiate resource flows but also influence the spatiotemporal dynamics of the construction process, thereby affecting efficiency, resource utilization, and safety outcomes. Multi-agent simulation (MAS) offers an in-depth simulation that elucidates resource flows and spatiotemporal shifts [4], allowing for the passive testing of various strategies [5]. It is only by integrating performers' subjective decision-making that the simulation can actively develop more adaptive strategies, thereby revealing novel insights into real-world construction processes.

At present, many construction decision issues are formulated as mixed-integer programming problems and pre-optimized through heuristic algorithms. These include adjusting scheduling plans [6], revising site layouts [7], and optimizing resource allocations [8]. Such approaches are generally apt for managers overseeing large-scale processes or for highly automated performers in stable environments. Examples include the multi-step production of components [9] or the development of wall building plans for unmanned aerial vehicles [10]. However, real-world construction processes involve numerous workers, each with their own subjective intentions, and they operate in environments filled with uncertainties. Consequently, performers must rapidly make autonomous decisions in response to their immediate surroundings.

Distributed strategy modeling and optimization can be effectively conducted using agent-based methods, which are further enhanced through the application of game theory [11] or Reinforcement Learning (RL). These approaches describe and analyze the system dynamics from a bottom-up perspective, optimizing the overall strategy by focusing on both individual and collaborative tactics [12]. RL is particularly beneficial in tackling decision-making challenges, uncertainty, and environmental feedback [13]. When integrated with MAS, RL gives rise to multiagent Reinforcement Learning (MARL). MARL is deemed highly suitable for the planning and control of tasks involving multiple agents [14], and it plays a crucial role in helping agents determine optimal decision policies [15].

The effective use of MARL necessitates the transformation of relevant problems into Markov Decision Process (MDP) frameworks [16]. MDP, an advanced form of Markov chains, incorporates actions and rewards, offering a mathematical framework for decision-making in scenarios characterized by both randomness and the influence of decision-makers' choices. MARL has been instrumental in modeling and optimizing construction action strategies, such as guiding robots to construct desired structures [17] and building linear barriers with filled bags [18]. In the realm of construction, MARL's primary application has been in planning scenarios [19], for instance, training agents for site layout optimization [20]. However, these studies predominantly focus on pre-planned solutions, rendering them less effective for real-time decision-making in ongoing construction activities. In the complex environment of practical construction, agents frequently encounter dynamically changing conditions, which present challenges like intelligent navigation and swift obstacle avoidance [21].

This challenge can be mitigated by employing real-time navigation for agents using Deep Reinforcement Learning (DRL), which has demonstrated effectiveness in addressing motion and decision-making challenges in complex and dynamic environments [22], including robot navigation in building interiors [23]. However, existing methods of real-time navigation are not entirely suitable for construction processes. In these settings, agents often spend a significant amount of time executing tasks after arriving at their target positions. As a result, their movements yield less frequent rewards compared to the existing research focused on multi-agent navigation, thus presenting unique challenges to Reinforcement Learning in construction contexts. More importantly, construction involves a series of specialized steps; therefore, focusing solely on movement decisions does not adequately address the comprehensive requirements of construction tasks.

Consequently, the objective of this study is to develop a Construction Markov Decision Process (CMDP) framework. This framework aims to convert fine-grained construction simulation into a MDP format and utilize Reinforcement Learning (RL) for optimizing agents' policies in dynamically evaluating and reaching construction targets. The CMDP framework comprises four primary components: a) a construction knowledge-based decision model, b) construction and physical state transition probabilities, c) construction state mapping and observation modification, and d) a hierarchical decision policy for target evaluation and reaching. To tackle the issue of sparse rewards in construction environments, the RL algorithm is applied using a two-stage training method. Finally, this framework has been developed on the Unity platform and its efficacy is demonstrated in a floor construction process, where it is compared to traditional construction strategies and heuristic methods.

The innovative aspects of this CMDP include the integration of construction knowledge into agent decision-making in various construction states, the standardization of decision inputs and outputs across different crew agents, and the implementation of a hierarchical policy model to effectively achieve construction goals. Furthermore, the introduction of a two-stage training method significantly expedites the policy training process for target-reaching objectives. This framework makes a substantial contribution by establishing a MDP tailored to the construction industry, enabling intelligent simulation of construction activities. It also identifies an interleaved construction strategy that significantly enhances efficiency in multiple dimensions of the construction process.

## 2. Related works

### 2.1. MARL in construction decision-making

Considering the inherently multi-participant nature of construction, MARL is deemed highly suitable for task planning and control in multi-agent scenarios [14]. RL methods are designed to identify optimal strategies for sequential decisions amidst uncertain rewards and future states [24]. In the context of construction, applying MARL necessitates converting relevant issues into MDP formats [16]. For example, Wei et al. [25] employed DRL to manage bridge component maintenance, encoding the state of bridge components as agent states within an MDP framework. Recently, RL has been applied to a variety of construction-related scenarios, including maintenance strategy planning for components [26], [27], resource allocation [28], schedule optimization [29], and machinery operation on construction sites [30]. Therefore, to effectively use RL in short-term construction processes, which involve modeling actors and considering resource flows, it is crucial to transform agents' spatial actions and decision-making related to target selection into an MDP framework.

Regarding the modeling and decision-making in contractors' behavior, researchers have developed multi-agent construction systems that generate low-level rules for independent climbing robots, ensuring the creation of desired structures [17]. In a similar vein, Soleymani et al. [18] introduced an autonomous system for robots to construct linear barriers. Focusing on specific construction tasks, Li and Zou [31] created an RL environment for robotic arms to execute tasks like transportation, picking, and installation of window panels. Although the motion paths in these studies are relatively straightforward, the technical challenges that construction agents face, such as intelligent navigation and reactive obstacle avoidance, remain unresolved [21].

In practical construction settings, multi-agent navigation methods are instrumental in planning the positioning and movement paths of equipment and workers on site [32]. For instance, Kim et al. [20] applied deep reinforcement learning to create realistic paths for an individual worker navigating a site to reach designated target locations. Similarly, Liu et al. [33] adapted rebar design into a path-planning challenge to ensure clash-free rebar layouts. However, Markov Decision Process (MDP) frameworks designed for site planning often involve static objectives and movement environments for agents. This characteristic renders them less adaptable to the dynamic and unpredictable environments typically encountered in construction processes.

*2.2. MARL in dynamic environments*

In dynamic environments, MARL methods have demonstrated success in computer and industrial sectors, particularly for real-time navigation and decision-making by agents [34]. For example, Semnani et al. [22] developed a hybrid algorithm that merges deep reinforcement learning with motion planning, specifically for multi-agent movement in highly dynamic environments. In the context of building environments, Nguyen et al. [23] utilized a combination of transfer learning and distributed MARL to address collision-free navigation challenges for groups of robots in building interiors. Moreover, intelligent navigation methods applied in such dynamic settings are also beneficial for assisting robots or drones with straightforward construction tasks. Zhang et al. [35] introduced a multi-robot path-planning framework designed for 3D aerial printing during a building project. In most related research, the agent's target is predetermined, enabling the training process to be streamlined through the analysis of the relationship between the target and the agent's state, such as the angle between the agent's orientation and the direction towards the target [36][37].

However, in actual construction practice, the targets for agents are frequently changing, and often there are several viable targets available at once. This presents a significant challenge in assigning fixed, deterministic targets to construction agents. The concept of reaching targets without pre-established goals has been investigated in recent studies. Xia et al. [38] introduced information entropy in the context of MARL to enhance the search range and stability of unmanned aerial vehicles (UAVs) when tracking multiple potential targets. Additionally, Chen et al. [39] developed an advanced MARL approach to coordinate multiple UAVs for real-time target tracking. In construction settings, agents must determine their current target based on their status and construction-specific knowledge. Current research in this area, particularly concerning the dynamic evaluation and selection of targets by agents, often overlooks the implications of target selection on construction behavior [36]. The primary focus tends to be on the cost of reaching the target [36] and potential collisions [40], rather than the broader consequences in the construction context.

On the other hand, proficient target perception is a crucial skill for multi-agent navigation in partially observable environments. Numerous studies employ a ray-based method to scan the environment and identify targets [41], thereby enhancing the agents' adaptability to real-world physical settings [42]. Benavidez and Jamshidi [43] implemented a system for robot navigation and obstacle avoidance, utilizing information derived from cameras. Zhang et al. [44] employed ray-cast sensors to facilitate cooperative task completion and navigation for UGVs and UAVs. In the context of construction, Zhang et al. [45] developed a hybrid approach combining task planning and search-based motion planning for an autonomous excavator. This excavator, equipped with LiDAR and camera sensors, calculates feasible paths for excavation tasks.

*2.3. Research problem identification*

Existing research highlights the lack of a Markov Decision Process (MDP) framework specifically tailored for fine-grained construction processes, which is crucial for training and simulating agents in subjective decision-making scenarios. While current real-time multi-agent navigation methods offer some reference points for this framework, three primary research challenges remain:

(1) The MDP for the construction process should integrate professional knowledge to accurately perceive construction states and make relevant decisions. Although some methods combine the task network of construction or manufacturing processes with agent decision-making to establish an MDP process [46] [47], the agent typically acts more as a scheduler or manager, rather than as an executor. The task states and resource flow within the execution environment, which must be comprehended through construction knowledge, are not adequately captured as observations in the MDP process.

(2) A dynamic policy model is essential for managing the varying types and quantities of targets to achieve higher-level construction goals. The construction behaviors of agents can profoundly and enduringly affect the environmental state. Existing methods typically assign targets using simplistic rules [22] [34], or optimize target selection by considering impacts prior to reaching them [48]. Furthermore, the distinct construction targets for different types of agents necessitate separate strategies, which could substantially extend the training duration.

(3) The training process for the strategy must account for the impact of sparse rewards in construction scenarios. In construction, agents spend a considerable amount of time performing tasks after reaching their target locations, leading to fewer and more sporadic rewards for their movements.

To address the abovementioned research problems, this paper introduces a CMDP framework for construction processes, incorporating embedded construction knowledge. This framework is capable of training agents' hierarchical policies to efficiently target high-value construction objectives. Furthermore, a two-stage training method utilizing reinforcement learning is implemented for training the agents' policies.

## 3. CMDP framework of construction process

### 3.1. Problem formulation

This research simulates the construction process, featuring multiple crew agents. In this simulation, agents concurrently navigate the work plane to identify and approach their targets, such as components, storage areas, and zones designated for construction activities. These activities require a specific duration to complete and can alter the physical state of the work plane. For instance, when an agent engages with a component to initiate a task, it generates an occupied region on the work plane, potentially hindering or decelerating the movement of other agents. This congestion is often a result of nearby materials, procedural measures, and safety considerations inherent in the actual construction of the components.

The construction targets for the agents are defined by the current state of the work surface, the overall construction progress, and the individual statuses of the agents. In real-world construction scenarios, workers depend on visual cues to discern different physical conditions and construction statuses on the work plane, often cluttered with obstacles and blockages. They apply their construction knowledge to identify targets and decide on appropriate construction actions. Subsequently, they strategize their movements to rapidly and safely reach these targets for action execution.

This description is formulated based on the following assumptions: (1) Workers within a single crew are treated as a collective agent in terms of actions and decision-making, reflecting the reality that crew members typically operate in close, collaborative units. (2) The movements and construction actions of the crew agents are modeled in two dimensions, which is suitable for the context of floor construction processes. (3) The efficiency of the crew is considered constant, unaffected by factors like fatigue, and the probabilities of state changes—stemming from movement, task progression, material acquisition, etc.—are treated as independent at each step. This approach ensures that subsequent states are determined exclusively by the current environmental state, in accordance with the principles of a Markov chain.

## 3.2. Construction Markov Decision Process

Given the defined problem and the existing preliminary research, the movement of crews on the work plane can be conceptualized as a dynamic challenge of perceiving and reaching targets on the work surface. In this context, agents are required to navigate the complexities of obstacles and congestion on the work plane, while prioritizing high-value targets.

The CMDP framework, depicted in Fig. 1, consists of four main components: a) A construction knowledge-based decision model; b) State transition probabilities; c) Mapping of construction states and modification of observations; and d) A hierarchical decision policy for evaluating and reaching targets.

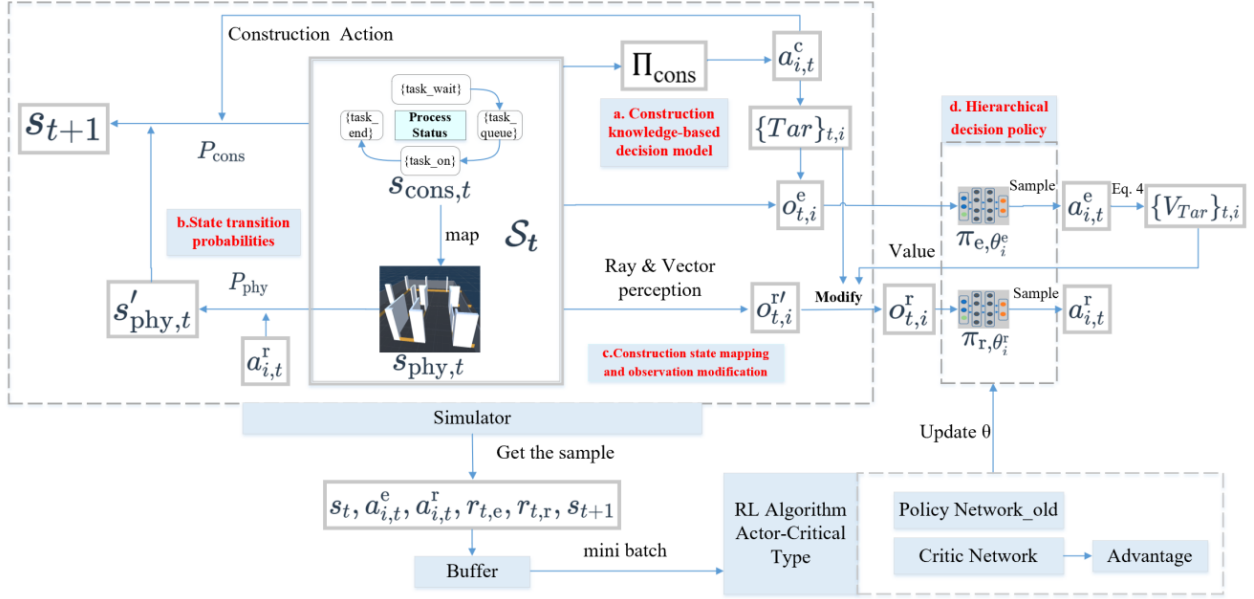

**Fig. 1.** The proposed CMDP framework

The proposed CMDP, as depicted in Fig. 1, is defined as the tuple: $< N, S, \{O_i\}_{i \in N}, \{A_i\}_{i \in N}, \{R_i\}_{i \in N}, P_{\text{phy}}, P_{\text{cons}}, \gamma >$, where N is the set of agents $i$, each $i$ holds a hierarchy policy and a construction knowledge decision model: target evaluating policy $\pi_{e,\theta_i^e}$, target reaching policy $\pi_{r,\theta_i^r}$, and construction knowledge decision model $\Pi_{\text{cons}}$. $S$ is the set of state consisting of physical state $S_{\text{phy}}$ and construction state $S_{\text{cons}}$. $\{O_i\}_{i \in N}$, $\{A_i\}_{i \in N}$, $\{R_i\}_{i \in N}$ are the observation, the action set and the reward of each agent. $\{O_i\}_{i \in N}$ includes the observation $\{o_{t,i}^e\}_{i \in N}$ for high-level target evaluation policy and the observation $\{o_{t,i}^r\}_{i \in N}$ for low-level target reaching policy. $\{A_i\}_{i \in N}$ includes construction action $\{A_i^c\}_{i \in N}$, target evaluating action $\{A_i^e\}_{i \in N}$, and target reaching action $\{A_i^r\}_{i \in N}$. $\{R_i\}_{i \in N}$ consists of rewards for target evaluation policy $\{R_{e,i}\}_{i \in N}$ and rewards for target reaching policy $\{R_{r,i}\}_{i \in N}$. $P_{\text{phy}}$ is the physical state transition probability used to calculate the next physical state based on agent reaching actions sampled from target reaching policy, primarily affected by agent physical attributes such as moving velocity $v$; $P_{\text{cons}}$ represents the construction state transition probabilities for evolving construction state according to the construction actions derived from construction knowledge decision $\Pi_{\text{cons}}$, mainly influenced by construction efficiency metrics like tasking efficiency.. Finally, $\gamma$ is the discount factor.

The CMDP execution process is shown in Fig. 1 and depicted as follows: For agent $i$, let the current time step be denoted as $t$, and there exists a state $s_t \in S$, where $s_t = s_{\text{phy},t} + s_{\text{cons},t}$. Initially, the construction decision $\Pi_{\text{cons}}$ is applied to obtain a construction action $a_{i,t}^c \in \{A_i^c\}$ based on $s_t$. If the agent needs to acquire targets or update target values at this step, $\Pi_{\text{cons}}$ outputs the current agent's target scope $\{Tar\}_{t,i}$. Following this, the agent's high-level target evaluation policy $\pi_{e,\theta_i^e}$ is activated, and an observation $o_{t,i}^e$ is collected from $s_t$ with reference to $\{Tar\}_{t,i}$. $\pi_{e,\theta_i^e}$ then generates an action $a_{i,t}^e \in \{A_i^e\}$, representing the relative target value $a_{i,t}^e \to \{V_{Tar}\}_{t,i}$ of targets. While holding $\{Tar\}_{t,i}$ and $\{V_{Tar}\}_{t,i}$, the agent collects observation $o_{t,i}^{r\prime}$ from $s_t$, modifies it to obtain $o_{t,i}^r$, and then inputs $o_{t,i}^r$ into its low-level target reaching policy $\pi_{r,\theta_i^r}$ that outputs a movement action $a_{i,t}^r \in \{A_i^r\}$.

Executing $a_{i,t}^r$ in state $s_t$ results in a physical state change $P_{\text{phy}}: s_{\text{phy},t} \times a_{i,t}^r \to s_{\text{phy},t}'$, followed by performing $a_{i,t}^c$ in the state $\{s_{\text{phy}}', s_{\text{cons}}\}$, which may result in changes in both construction and physical states: $P_{\text{cons}}: s_{\text{phy},t}' \times s_{\text{cons},t} \times a_{i,t}^c \to s_{\text{cons},t+1}, s_{\text{phy},t+1} \to s_{t+1}$. The state $s_{t+1}$ produces rewards for the evaluation and reaching policies: $r_{e,t} = R_e(s_{t+1})$, $r_{r,t} = R_r(s_{t+1})$. The agent's goal is to optimize its policies $\pi_{e,\theta_i^e}$ and $\pi_{r,\theta_i^r}$ to maximize the cumulative reward $\sum_{t=0}^{T} \gamma^t r_{e,t} + \sum_{t=0}^{T} \gamma^t r_{r,t}$ over time. The expected future discounted reward from executing policies with parameters $\theta_i^e$ and $\theta_i^r$ is $E_{S\sim P, a_e \sim \pi_{e,\theta^e}}[\sum_{t=0}^{T} \gamma^t r_{e,t}] + E_{S\sim P, a_r \sim \pi_{r,\theta^r}}[\sum_{t=0}^{T} \gamma^t r_{r,t}]$.

*3.3. Construction knowledge-based decision model*

Construction knowledge encompasses the entire range of operational procedures for crew work at construction sites, including equipment preparation, material handling for tasks, and task execution. The use of construction knowledge as the decision model $\Pi_{\text{cons}}$ is intended to guide the derivation of construction actions $a_{i,t}^c \in \{A_i^c\}$ or target range $\{Tar\}_{t,i}$ under specific states, and to determine the corresponding state transitions. In the construction process, where strict adherence to predefined processes and patterns is essential, the decision model $\Pi_{\text{cons}}$ based on construction knowledge, employs a "decision table," as illustrated in Table 1. This approach is chosen to avoid the use of neural networks, which could introduce instability and behaviors inconsistent with established construction logic.

For example, when the construction status of the environment indicates the existence of an executable task and the agent's status indicates that it is not currently performing a task and has sufficient materials, the output of $\Pi_{\text{cons}}$ can determine the agent's target type $\{Tar\}_{t,i}$, aligning with the component linked to the relevant executable task. Upon the agent's arrival in proximity to such a component, it can commence the tasking action. This construction knowledge-based decision model, tailored to the floor construction process, is illustrated in Table 1.

**Table 1.** The construction knowledge-based decision model

| Code | Name | Trigger state | State transition |
|---|---|---|---|
| $a^{c1}$ | Initiate task | Reach task target and equipment sufficient | Task area generated in $s_{\text{phy}}$, tasks flow in $s_{\text{cons}}$. Agent state changes to tasking. |
| $a^{c2}$ | Execute task | Agent in tasking state and equipment and material are sufficient | Task process promoting in $s_{\text{cons}}$ according to $P_{\text{cons}}$. |
| $a^{c3}$ | Pause task | Agent in tasking state while equipment or material are going insufficient | Agent state changes to target reaching, its target scope $\{Tar\}_{t,i}$ is set to relate storage or equipment. |
| $a^{c4}$ | Resume task | Reaching task target that was initiated | Agent state changes to tasking. |
| $a^{c5}$ | Fetch materials | Reaching storage target with sufficient stock. | Agent state changes to fetching materials, reducing storage stocks and increasing agent's load in $s_{\text{cons}}$ |
| | | | according to $P_{\text{cons}}$. Updating the congestion index of storage in $s_{\text{phy}}$. |
| $a^{c6}$ | Fetching completed | Agent in fetching state and about to maximum load | Agent state changes to target reaching, target scope $\{Tar\}_{t,i}$ is set to its paused task. |
| $a^{c7}$ | Request crane for material | Agent in fetching state while storage stock is running out | Adding agent's request to cane in $s_{\text{cons}}$, agent state changes to waiting until the lifted material arrives. |
| $a^{c8}$ | Request crane for task | Started a task that require carne assistant, and material sufficient | Adding agent's request to cane in $s_{\text{cons}}$ agent state changes to waiting until the carne assistant arrives. |
| $a^{c9}$ | Complete task | Agent in tasking state and task process about to compete | Deregister task area in $s_{\text{phy}}$, drive tasks flow in $s_{\text{cons}}$, and agent state changes to target reaching. |
| $a^{c10}$ | Acquire target | Agent in target reaching state while holds no target scope and paused target | Agent's target scope is set to components with executable tasks of its type. If no tasks of the agent's type exist in the task queue and task wait, target is set to the outlet, and state changes to deregistered. |

## 3.4. Construction and physical sate transition probabilities

This CMDP consists of two types of transition functions: $P_{\text{cons}}$ and $P_{\text{phy}}$. $P_{\text{phy}}$ describes the transition of physical states caused by agent movement, primarily influenced by the agent's forward velocity $v_i^f$, lateral velocity $v_i^l$, and turning rate $v_i^{tu}$. $P_{\text{cons}}$ implements the actual construction state changes represented by task progress and material reserves, primarily influenced by the agent's task efficiency $e_i^t$ and material acquisition efficiency $e_i^m$. All the simulation uncertainties are defined in $P_{\text{cons}}$ and $P_{\text{phy}}$, Where the velocity ratio in $P_{\text{phy}}$ will be determined by agent's policy which is a sample from neural network, the efficiency in $P_{\text{cons}}$ will be a sample from distribution established from real data. Besides, to account for real-world interference and blockages in construction scenarios, both state transition probabilities will consider various efficiency and obstruction factors.

Regarding $P_{\text{phy}}$, agents' movement velocities are reduced by various congested areas, such as tasking areas and storage areas. The current maximum velocities $v_{i,t}$ are calculated as follows:

$$v_{i,t} = \max(c \times I_t^{\text{area}} \times v_i, v_{i,\min}) \quad (1)$$

where $v_i$ is agent nominal velocity, which can be either of $v_i^f$, or $v_i^l$, or $v_i^{tu}$, and $v_{i,\min}$ is the lowest velocity among these three to ensure that the agent can escape from congested areas. $I_t^{\text{area}} \in [0,1]$ represents agent's current congestion index coursed by the congestion areas it is located in,

$I_t^{area}$ value in storage area will be decreased with the material reserves, and in other areas, it is determined by task type or structural zone and is set as a fixed value. $c \in [0,1]$ is the deceleration index, which is smaller than one when agent carrying materials.

Regarding $P_{cons}$, the agent's task efficiency $e_i^t$ and material acquisition efficiency $e_i^m$ are affected when other agents invade the same area where it is performing action, that is:

$$e_{i,t} = \max\left(0, e_i \times \left(1 - n \times I_t^{agent}\right)\right) \qquad (2)$$

where $e_i$ is the agent's nominal action efficiency, which can be either $e_i^t$ or $e_i^m$. $I_t^{agent}$ is the inefficiency index caused by the invasion of other agents, and $n$ is the number of invading agents.

### 3.5. Construction state mapping and the observation modification

Agent observations depict how the agent perceives its surrounding environment and guide the agent in achieving its action objectives. In navigation problems, ray perception is often used to simulate how humans or machines collect surrounding observations in the real world, and it is highly effective for agents to perceive targets as well as obstacles. However, the agent's target in this framework is highly related to the construction status, which cannot be directly perceived by ray casting in the physical environment.

The construction state $S_{cons}$ in this framework is defined by the task flows among {task_wait}, {task_queue}, {task_on}, and {task_end}, representing four task pools where tasks are driven by task precedence relationships [49]. As all construction tasks are associated with components and executed around them, this framework projects the tasks in {task_queue} and {task_on} task pools to the component as component's performing task and executable task attributes in real-time. Consequently, agents can use ray observations to perceive components' corresponding attributes to know which tasks are waiting for performing and where they are.

Due to the varied target types of different crew agent types, directly using target scope as input for agent policy would significantly increase the diversity of policies among different agent types and lead to prolonged training times. Thus, this CMDP modifies agent observation directly with reference to $\{Tar\}_{t,i}$ and $\{V_{Tar}\}_{t,i}$. $\{Tar\}_{t,i}$ modifies agents' observation contains its current feasible target or not, $\{V_{Tar}\}_{t,i}$ make agent know how valuable the observed target is, the higher value target promising a high reward when arrive. So that different agents can share the same decision policy for target reaching. Specifically, the modification replaces the priority value of ray perception outcomes with the target's priority values:

$$\forall \text{ray}_j \in o^r_{t,i}, V_j = \begin{cases} V_{ray_j}, ray_j \in \{Tar\}_{t,i} \\ 0, ray_j \notin \{Tar\}_{t,i} \end{cases} \quad (3)$$

where $ray_j$ represents the $j$ th ray perception outcome, $V_j$ is the priority value of $ray_j$, $\{Tar\}_{t,i}$ stands for the current target scope of the agent $i$, and $V_{ray_j}$ signifies the priority value of the observed target, which is obtained from Target evaluation policy.

*3.6. Hierarchical decision policy for target evaluation and reaching*

The hierarchical decision policy of this CMDP comprises a high-level target evaluation policy $\pi_{e, \theta^e_i}$ and a lower-level target-reaching policy $\pi_{r, \theta^r_i}$, shown as Fig. 2. When an agent acquires its target range or after a certain number of steps without reaching any target, the high-level policy $\pi_{e, \theta^e_i}$ is activated to collect its observations from the current state with reference to each target in $\{Tar\}_{t,i}$ and output relative priority values for the target range. These outputs then serve as the basis for modifying observations in the lower-level policy $\pi_{r, \theta^r_i}$, thereby integrating high-level construction policies trained in the high-level policy with the execution in the low-level policy. The agent's lower-level policy is activated at each time step, except for tasking, waiting, and deregistered states. It collects observation data and then outputs movement actions to swiftly reach high-value targets. If an agent holds no target, it should maneuver flexibly to avoid collisions.

Both the high-level and low-level policies are modeled using neural networks, with their outputs being samples from the network outputs at each decision step. This enables agents to rapidly determine actions based on the current environmental conditions, continuously exploring environmental rewards for gradual strategy optimization. In a multi-agent context, the decisions of other agents introduce elements of uncertainty, thus necessitating centralized training to ensure stable policy iteration.

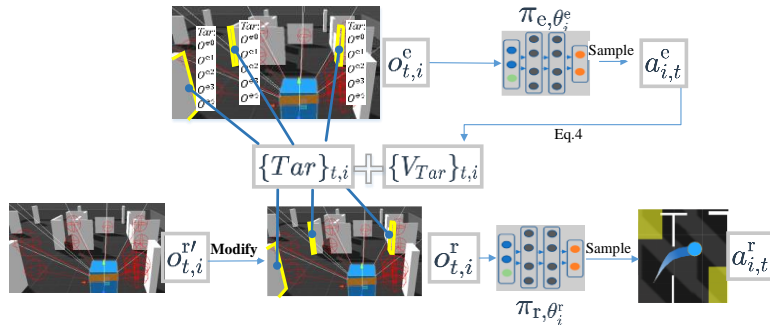

**Fig. 2.** The structure of hierarchy policy

*3.6.1. Target evaluation policy*

The training objectives of the high-level target evaluation policy are to achieve construction goals, which can be improving efficiency, shortening construction duration, enhancing safety, and equalizing labor load. For the sake of effective training, this policy will be designed specifically for the construction objectives of improving efficiency and shortening the duration.

**Observation space of evaluation policy**

The observation collection for the evaluation policy involves gathering data related to potential efficiency impact within a certain range of the agent. While the distance between the agent and the target seems to be an obvious factor to collect, construction scenarios introduce intricate obstructions and conflicts to agent movement. Additionally, reaching the task target initiates a congestion area, leading to profound changes in the environment state. Furthermore, during tasking, the existence of logical and spatial constraints affects other agents' task scope, influencing the initiation of subsequent tasks. Therefore, target evaluation requires the consideration of a broader array of factors.

This framework primarily considers influencing factors from the perspective of improving construction efficiency: the time costs for the agent to reach the target, the logical constraint impact of the target, the spatial constraint impact of target, and the spatial obstruction impact of target arrival on agent movement. To ensure a consistent observation space size, each time the evaluation policy's observation will consider all potential targets throughout the entire game episode. The observation space $O^e$ that $\pi_{e,\theta_i^e}$ needs to observe includes:

$O^{e0}$ : Object attributes such as its type, task pool it current belongs to, its position, its congestion area scale, its expected action time and so forth;

$O^{e1}$ : weather the object is agent's current target;

$O^{e2}$ : the distance between the agent and the object;

$O^{e3}$ : the total number of subsequent tasks for the object;

$O^{e4}$ : the number of tasks in the "queue" and "wait" pools will have space conflict with this object initiation.

**Action space of evaluation policy**

As the number of targets may vary in each round of target evaluation, this framework introduces a novel approach by depicting the action output of the evaluation policy as the most desired target position. The action space $A^e$ is designed with two continuous values, $a^{e1}, a^{e2} \in$

$[-1,1]$. Here, $a^{e1}, a^{e2}$ represent the normalized values of the $x$ and $y$ coordinates of the desired position, with the target's value being inversely proportional to the distance from the desired position. This definition of action outputs cleverly addresses the challenge of evaluating targets with varying types and quantities while maintaining a fixed output size. Moreover, using optimal positions as outputs enhances the training and utilization of the policy's ability to handle spatial layout in the construction process. The calculation for target priorities $V_{Tar}$ is as follows:

$$V_{Tar} = 1 - \sqrt{\left(a^{e}1 - \frac{x_{Tar}}{x_{max}}\right)^2 + \left(a^{e}2 - \frac{y_{Tar}}{y_{max}}\right)^2} \tag{4}$$

where $x_{Tar}$ and $y_{Tar}$ are the $x$ and $y$ coordinates of target position, $x_{max}$ and $y_{max}$ represents the maximum coordinate distances of the work plane.

**Reward design of evaluation policy**

In this framework, the impacts resulting from the evaluation policy take time to manifest and persist, as the execution period of the target and its subsequent targets may both affect construction efficiency. Consequently, it is necessary to monitor any factors related to construction efficiency from the simulation system to provide long-term rewards or penalties to target selection agents. The reward function is designed as follows:

$$R_e = R_{e,idel\_pre} + R_{e,idel\_area} + R_{e,efficiency} + R_{e,episode} + R_{e,path} \tag{5}$$

where $R_{e,idel\_pre}$ represents the penalty incurred when agents remain idle due to the non-initiation of the precedence target; $R_{e,idel\_area}$ represents the penalty incurred when the initiated task area causes space constraints, leading to the idleness of other agents; $R_{e,efficiency}$ is the reward related to the current construction resource utilization efficiency, where having fewer idle agents (agents with no ongoing construction actions and no targets) results in higher efficiency utilization;

$R_{e,episode}$ is the reward linked to the duration of the current game episode; $R_{e,path}$ is the reward associated with the total movement steps of the agents in the current game episode.

*3.6.2. Target reaching policy*

Target-reaching policy controls agent movement to efficiently reach higher-value targets as quickly as possible. This policy is triggered at each step unless the agent is in a state of tasking, material acquisition, waiting, or deregistration. Its purpose is to guide the agent efficiently toward high-value targets while avoiding obstacles.

**Observation space of reaching policy**

The observation model for the target-reaching policy is designed to simulate the visual perception of construction crews. This allows agents to perceive high-value targets, surrounding obstacles, and congestion areas. The agent's ray perception model is depicted in Fig. 3 and consists of two parts: frontal rays of sight and surrounding rays for obstacle detection. The *n* frontal rays extend horizontally from the front of the agent to the maximum sight distance, with their directions constrained within the agent's field of view. Surrounding detection involves *m* rays that are evenly projected from all sides of the agent, including the back, at a downward angle.

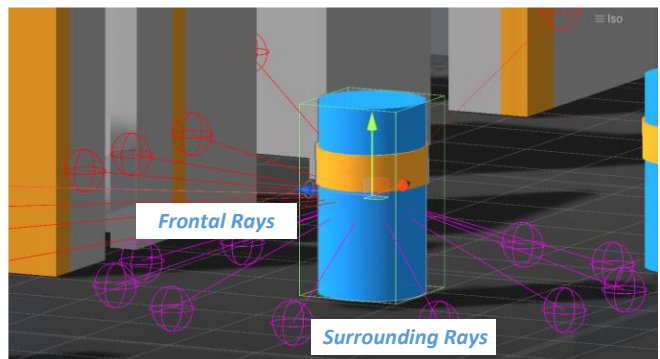

Fig. 3. Ray perception model of crew agent

Each ray, upon detecting an object and being modified with Eq.3, provides four values: the relative length of the ray to the object, the angle at which the ray was emitted, the numerical identifier of the detected object's type, and the priority value of the detected object. Additionally, the strategy requires observing the agent's current movement direction and velocity magnitude in the x and y directions to adjust the size and direction of movement in the strategy output. Therefore, the observation space of this strategy includes:

$$O^r = (n + m) \times 4 + v_x + v_y + |v|$$

**Action Space of reaching policy**

The movement of agents on the work surface can be considered two-dimensional, and its action space is composed of $A^r = \{v_i^f, v_i^l, v_i^{tu}\}$. $v_i^f$ represents the nominal value of the agent's forward movement velocity, and for safety reasons, the crew agents are unable to move backward. $v_i^l$ denotes the nominal value of the agent's lateral movement velocity, which is an important way for workers to avoid obstacles on the site. $v_i^{tu}$ represents the nominal value of the agent's turning rate.

To expedite training, the continuous action space encountered in actual construction is suitably discretized without a significant loss of accuracy. The following discretization is applied:

$$v_{i,t}^f \in [0, 0.3v_i^f, 0.8v_i^f, v_i^f];$$

$$v_{i,t}^l \in [-v_i^l, 0, v_i^l];$$

$$v_{i,t}^l \in [-v_i^{tu}, 0, v_i^{tu}].$$

**Reward design of reaching policy**

Agents need to minimize their entry into congestion areas and avoid collisions to reach currently observed high-value targets as quickly as possible. Therefore, the reward function of the target-reaching model in this framework is designed as follows:

$$R_r = \eta \times R_{r,\text{colision}} + R_{r,\text{reach}} + R_{r,\text{time}} \qquad (6)$$

where $R_{r,\text{colision}}$ is the punishment when agent enters congestion area or collides with obstacle,

and $\eta$ is the collision punishment index of different objects. Collision with wall, storage area are small $\eta_b$ index to allow agents to pass through barriers. Collision with other agents or task areas has a middle $\eta_a$ index because it causes congestion and slows down the movement of itself.

$R_{r,reach}$ is the reward when the agent reaches a target, it is determined by the target priority value:

$$R_{r,reach} = k \times V_{Tar} \tag{7}$$

where $k$ is the priority value scale, with a higher value resulting a stronger induction of high value targets. $R_{r,time}$ is a minor penalty imposed when the agent has a target to reach, it encourages agent to reach target as quickly as possible.

## 4. Reinforcement learning algorithm application and two stage training

### 4.1. Multi-agent proximal policy optimization algorithm application

The Actor-Critic approach of the reinforcement learning algorithm proposed in [50] involves centralized policy learning with decentralized policy execution that prompt training process. It also enables the aggregation of all agents' experiences into the entire agent group or cluster, aligning all agents' efforts towards a common goal. This is well-suited for training multi-agent execution strategies in construction scenarios where all agents share a common construction objective and operate within a fully cooperative framework. Moreover, this CMDP utilizes the construction knowledge-based decision model to harmonize the observations and action spaces for different crews with varying target types and target determination logics, so that the Actor-Critic method for centralized training while decentralized execution can be applied.

This paper employs the MAPPO algorithm for optimizing multi-agent movement in construction contexts. PPO (Proximal Policy Optimization) [51] is a gradient ascent policy optimization algorithm, belonging to the Actor-Critic family. PPO is suitable for both continuous and discrete action spaces, demonstrating success in various experiments. The MAPPO (Multi-Agent Proximal Policy Optimization) algorithm extends PPO to multi-agent reinforcement learning, enabling training of multiple agents in a collaborative setting using a shared value function to estimate individual agents' state values.

In this framework, the MAPPO algorithm is responsible for updating and maintaining the two-tiered policy networks for each agent, based on the expected reward $E_{S \sim P, a_e \sim \pi_{e, \theta^e}}\left[\sum_{t=0}^{T} \gamma^t r_{e,t}\right]$ and $E_{S \sim P, a_r \sim \pi_{r, \theta^r}}\left[\sum_{t=0}^{T} \gamma^t r_{r,t}\right]$, MAPPO optimizes the following objective function:

$$L^{\text{CLIP}}(\theta) = \mathrm{E}\left[\min\left(r_t(\theta)\widehat{\mathcal{A}}_t, \text{clip}(r_t(\theta), 1-\epsilon, 1+\epsilon)\widehat{\mathcal{A}}_t\right)\right] \tag{8}$$

where $r_t(\theta) = \frac{\pi_\theta(a_t|s_t)}{\pi_{\theta_{\text{old}}}(a_t|s_t)}$ denotes a probability ratio between new and old policies, $\theta$ represents two policies parameters $\theta^e$ and $\theta^r$ respectively, $\epsilon$ is a hyperparameter which defines the clip range, the clip function limits the update range to $[1-\epsilon, 1-\epsilon]$. $\widehat{\mathcal{A}}_t$ denotes the estimation of the advantage function at time $t$. By clipping the probability ratio, a too large policy update is restricted which results in robust learning performance. This CMDP applies MAPPO algorithm as follows.

**Algorithm:** MAPPO in CMDP

```
Initialize the neural network  π_{e,θ^e}  and  π_{r,θ^r};
Set maximize number of episodes  K;
While current episode ≤ K;
    Rest training environment to initial state  S_0;
    For each agent  i  in scene do
        Get current state  s_{t,i}  Get current construction action and targets using  Π_{cons}: a^c_{i,t}, {Tar}_{t,i} ←
        Π_{cons} × s_{t,i};
        Sample an evaluation action using  π_{e,θ^e}: a^e_{i,t} ← π_{e,θ^e} × s_{t,i} × {Tar}_{t,i}  Get target values using Eq.4:
        {V_{Tar}}_{t,i} ← a^e_{i,t} × {o^e_{t,i,j}}_{j=1,2,3,4} ;
        Get the  π_{r,θ^r}  observation  o^{r'}_{t,i}  and modify it using Eq.3:  o^r_{t,i} ← o^{r'}_{t,i} ;
        Sample a reaching action using  π_{r,θ^r}: a^e_{i,t} ← π_{r,θ^r} × o^r_{t,i};
        Simulate the physical state by  P_{phy}: s'_{phy,t} ← s_{phy,t} × {a^r_t}  Simulate the next step state by  P_{cons}: s_{t+1} ←
        s'_{phy,t} × s_{cons,t} × {a^c_t} ;
        Collect set of trajectories of each policy  D_e  and  D_r  for  T  time steps;
        Collect reward  r_e  and  r_r  for  T  time steps;
        Compute the advantage estimate  Â_1 …. Â_T;
    End For
    Optimize surrogate loss with minibatch for  K  episodes;
Update policies by maximizing their objective  L^{CLIP}(θ^e)  and  L^{CLIP}(θ^r)  function:  θ_{old} ← θ.
```

### 4.2. Two stage training

As agents in this CMDP are not assigned targets beforehand, the typical approach of setting movement rewards for agents before reaching their targets, common in navigation papers, is not applicable. Moreover, in construction scenarios, agents often spend more time on tasks than reaching targets. This leads to sparser navigation rewards in CMDP, impeding the reinforcement learning training convergence. To address this, the framework employs a two-stage reinforcement strategy to accelerate the training of agent hierarchy policies in construction scenarios.

The first stage focuses on training the agent's target-reaching policy, omitting task durations, crane requests, and storage stock consumption to boost reward density. This enables agents to promptly decide on the next target after reaching the current one. In the second stage, all excluded elements are reintroduced to train the agent's hierarchy policy in a complete CMDP process.

This two-stage approach rapidly improves the agent's target-reaching policy in the first stage, enhancing agile obstacle avoidance and swift arrival at high-value targets. These abilities accelerate the training of target-reaching policies in complex task areas and offer efficient execution for high-level policies in the second stage with sparse rewards.

5. Case study and result

To validate the effectiveness of the proposed reinforcement learning framework for construction scenarios, this study utilizes the floor construction of prefabricated structures from a previous investigation [5] as a case study.

The case study encompasses five different types of construction crew agents, totaling seven in number: one joint clean crew (JC), one hoist and install crew (HI), one grouting crew (G), two reinforcing crews (R), two frame-working crews (F), and a tower crane. These agents and resources are tasked with constructing 36 components through 87 distinct tasks. Detailed information about the task types and their spatial characteristics is provided in Table 2, while the agents and their spatiotemporal attributes are summarized in Table 3. The data for these tables were derived from actual engineering projects and recent construction video analyses. In line with research on human or humanoid robot simulation, it is noted that agents' turning rates typically do not exceed 45°/s [52]. Additionally, referencing bridge design literature, pedestrians' lateral acceleration is recommended to be capped at 0.1m/s² [53]. Considering that in compact construction scenarios, excessively high turning rates may lead to missing entry points between components, we set the turning rate $v_i^{tu}$ = 30°/s, and lateral movement speed $v_i^l$ = 0.1m/s.

This CMDP is developed using the Unity platform, Unity ML-Agents Toolkit [54], after importing the geometry model and the task network of this case study, setting the simulation and training parameters, the CMDP game of this case is established (Fig. 4). It will train the crew agents to reach and finish all the tasks as soon as possible. The developed CMDP in Unity is simulated and trained in a computer with 13[th] Gen Intel® Core i9-13900K (3.00 GHz), 64GB memory, with NVIDIA GeForce RTX 4090, and Windows 11 Professional Edition 64-bit operating system.

Table 2. The task types and attributes in this case study

| Task Name | Required agents | Area distance (m) | Congestion index | Material storage | Material acquire efficiency |
|---|---|---|---|---|---|
| Joint cutting and Cleaning (JC) | JCC | 1.2 | 0.9 | | |
| Hoisting and Installation PCs (HI) | TC, HIC | 2.5 | 0.4 | Struts | 1min |
| Grouting (G) | GC | 1.1 | 0.7 | | |
| Reinforcing (R) | RC | 1.5 | 0.7 | Rebars | $0.1m^3$/min |
| Form-working (F) | FC | 2.0 | 0.5 | Templates | $5m^2$/min |

Table 3. The composition and the time-space attributes of the crew agents.

| Agents | Work efficient /Duration (expected value) | Variance value | Workers and Equipment | Occupation area ($m^2$) | Velocity (m/s) |
|---|---|---|---|---|---|
| JCC | 11.8 min | 1.2 min | 3* | 4 | 0.4 |
| HIC | 11.5 min | 0.6 min | 5 | 5 | 0.3 |
| GC | 5 min | 0.2 min | 1* | 1.5 | 0.6 |
| RC | $0.021m^3$/min | $0.007m^3$/min | 2 | 2.4 | 0.7 |
| FC | $0.495m^2$/min | $0.134m^2$/min | 2 | 3.2 | 0.5 |

Note: * means this crew is equipped with equipment

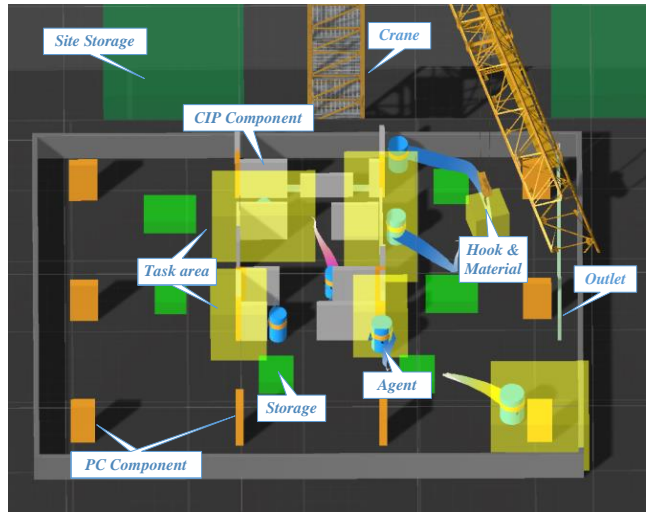

**Fig. 4.** The CMDP framework developed for this case study using Unity

*5.1. Parameters setting and initial state*

In this case study, the settings of this CMDP were as follows,

Simulation setting: The time step length was 1 second, agent's ray perception model is $n =12$ frontal rays and $m = 16$ surrounding rays.

Initial task environment status: All tasks are not started.

Initial work plane environment: The reserves of all storages are zero.

Congestion parameters setting: $c_{RC} = 0.9 > c_{HIC} = 0.8 > c_{FC} = 0.7$ are the deceleration index when carrying materials of agent RC, HIC and FC irrespectively. A higher $c$ value indicates a smaller deceleration effect. The deceleration is primarily determined by the material properties carried by the agents. The reinforcing steel agent carries small-volume steel bars, the HIC agent carries support poles with a larger volume, and the FC agent carries formwork, which is the largest and heaviest material. Therefore, there is a relationship $c_{RC} > c_{HIC} > c_{FC}$, and the specific values are obtained from video data. $I_t^{agent}$ is the agent inefficiency index to entered tasking areas, as defined in Eq. 2. Its value is connected to the agent's occupied area, and each agents' $I_t^{agent}$ is : $I_t^{JCC} = 0.4$, $I_t^{HIC} = 0.5$, $I_t^{GC} = 0.15$, $I_t^{RC} = 0.24$, $I_t^{FC} = 0.32$.

*5.2. Two stage training and results*

In the initial stage of training, the task duration is set to zero, with unrestricted stockpile reserves and no limitations on crane assistance for agents. This phase focuses on honing the agents' decision-making abilities regarding target attainment within the construction work plane. The rewards for the target-reaching policy during this stage are illustrated in Fig. 5. Observations reveal a rapid increase in the agents' rewards during the first million training steps. This is followed by a plateau, where rewards reach their peak and stabilize at around six million steps. Such a trend signifies that the agents' lower-level strategy effectively overcomes obstacles and blockages to promptly reach high-value targets, thereby establishing a solid foundation for the subsequent stage of training.

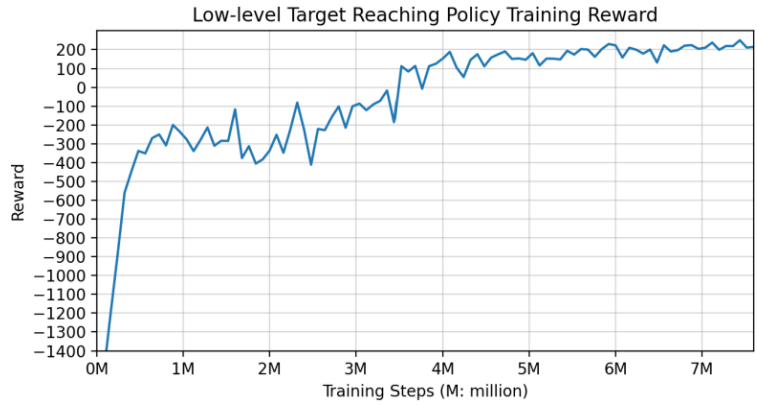

**Fig. 5.** Training rewards for the target-reaching policy in first stage

The training then progresses to the second stage, where all simplified settings in the simulation environment are removed, and the training algorithm's hyperparameters are fine-tuned to better suit the sparser reward MDP process. The outcomes of the target-reaching training in this stage are shown in Fig. 6. An initial decline in rewards is observed as the low-level strategy training commences at the second stage (around 7.6 million training steps), attributable to the introduction of task duration. This is followed by a slow but steady increase in rewards, which eventually stabilize at around 35 million steps. This stabilization point coincides with the evaluation strategy reaching approximately 15,000 training steps. It is important to note the asynchronous nature of the training steps between the high and low-level policies. While the target-reaching steps stabilize at 35 million, the target evaluation policy has undergone approximately 150,000 training steps.

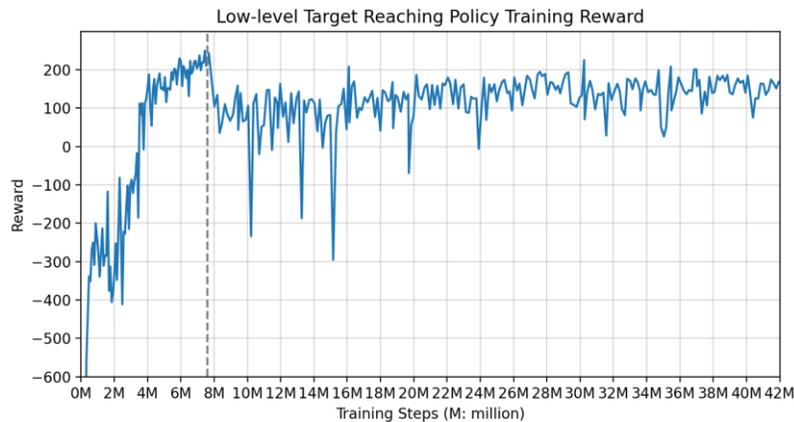

**Fig. 6.** Two stages training rewards for the target-reaching policy

The training outcomes for the target evaluation policy across both stages are depicted in Fig. 7. At the onset of the second stage (around training step 15,000), a sharp decrease in rewards is observed. However, as the policy adjusts to the prevailing environmental conditions, there is a swift increase in rewards, noticeable from step 65,000 to step 210,000. This is followed by a consistent rise in rewards from step 210,000 to step 520,000, culminating in a stable reward phase between half a million and one million steps. The stabilization of the target-reaching policy by the time the target evaluation policy has reached 150,000 training steps suggests that the high-level policy is effectively optimizing the spatiotemporal dynamics of task execution. This optimization leads to enhanced rewards and the establishment of a relatively stable strategy.

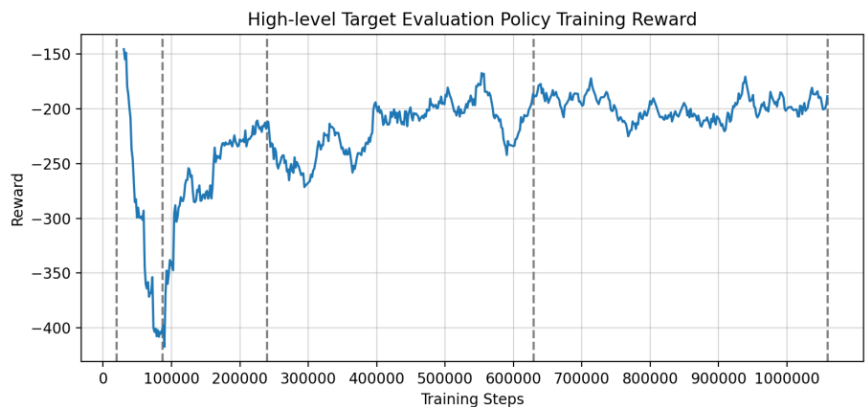

**Fig. 7.** Two stages training rewards for the target evaluation policy

With a trained hierarchical policy in place, agents within this CMDP framework achieve an intelligent simulation of the case study, with a portion of the simulation status showcased in Fig. 8. Following the simulation, Gantt charts for agents and tasks are automatically generated, as seen in Figs. 9 and 10, effectively reproducing the entire construction process, inclusive of varied agent statuses and task sequencing. In Fig. 9, four categories of agent actions are visually differentiated: idle (gray), navigating (yellow), task execution (red), and material acquisition or crane waiting (green, though many instances are too brief to be distinctly visible). Fig. 10 displays all 86 tasks completed in the simulation, each represented by a bar; the bar's left end marks the task's commencement, while the right end signifies its completion. The tasks are arranged vertically in ascending order of their start times. Due to the detailed nature of the simulation and the multitude of tasks, the display of individual task numbers is excluded for clarity.

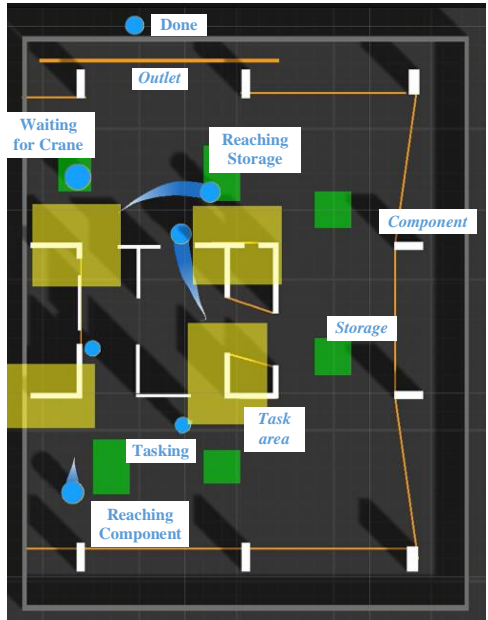

**Fig. 8.** The simulation status illustration of CMDP in this case study

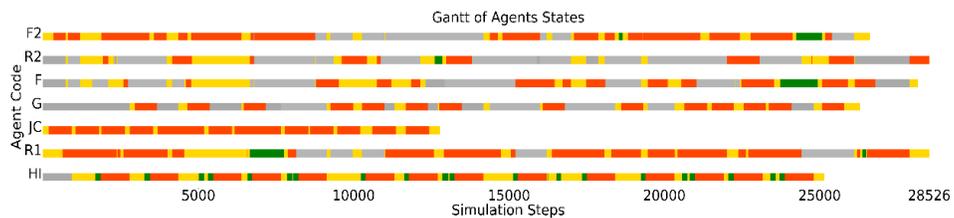

**Fig. 9.** The agents' Gantt chart of the trained hierarchy policy's simulation result

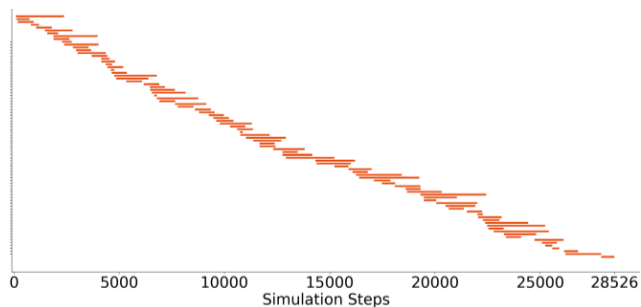

**Fig. 10.** The tasks Gantt chart of trained hierarchy policy's simulation result

Furthermore, the evaluation policy reveals novel construction patterns, which are characterized as an innovative construction strategy. Fig. 11 provides an analysis of the trajectories and execution sequences of components by the HIC agent in the simulation. The task objects for the HIC agent are prefabricated components, indicated in yellow, with their execution order marked in proximity. The yellow lines represent the HIC agent's trajectory, with narrower lines signifying earlier trajectories. Notably, the prefabricated component situated at the center functions as a preliminary task component for the adjacent cast-in-place components, thus warranting a higher precedence in the execution order.

Notably, the evaluation policy autonomously developed a strategy for sequential block-based construction, progressing from part 1 to part 5 as illustrated in Fig. 11. This strategy involves alternating between tasks on peripheral and central components within each block. Such an approach facilitates initial avoidance of tasking in the central area to reduce spatial constraints and strategically aims to complete certain HI tasks on central components promptly. This is done to remove pre-existing constraints for surrounding cast-in-place components, thereby enhancing overall construction efficiency.

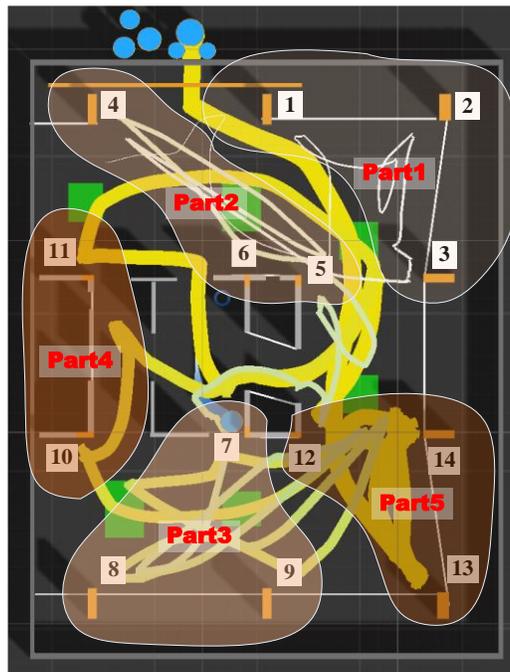

**Fig. 11.** The trajectory and execution pattern of HIC with the trained hierarchy policy

### 5.3. Comparison to other optimization methods

To validate the efficacy of the trained construction strategies, a comparative analysis was conducted between the high-level policy and two alternative approaches: a target priority calculation method and a genetic algorithm (GA) method.

The target priority method converts construction experience into actionable policies, enabling agents to identify the optimal target based on their current status. This method parallels strategies used in multi-agent navigation problems, where factors like distance [34] [55] or collision condition status [56] between the agent and target are considered for priority determination. In the context of construction, prioritization hinges on target attributes such as successors, conflicting tasks, and other relevant factors. The priority calculation follows previous study [5] as:

$$V'_{Tar} = \beta_1 \times \frac{Suc_{Tar}}{\max\{Suc\}_{\{Tar\}_{t,i}}} + \beta_2 \times \left(1 - \frac{Dis_{Tar}}{\max\{Dis\}_{\{Tar\}_{t,i}}}\right) + \beta_3 \times \frac{Spa_{Tar}}{\max\{Spa\}_{\{Tar\}_{t,i}}} \quad (9)$$

where $Suc_{Tar}$ is the number of successor tasks of target task $Tar$, $Dis_{Tar}$ is the distance between the target and the agent, and $Spa_{Tar}$ is the number of space conflicts between the target and other tasks in the task queue {task_que}. In this case study, $\beta_1$, $\beta_2$ and $\beta_3$ are set to 1, 0.5, and -1, respectively.

The GA is widely employed in engineering schedule optimization and is also applied here to provide a comparison with the high-level policy. In light of the uncertainties inherent in movements controlled by agent neural networks, GA genes are defined based on task priorities, as outlined in reference [57]. Task constraints in this context include spatial, resource, and prerequisite requirements, while the adaptability function focuses on minimizing the agent simulation duration and reducing the number of idle agent steps.

By employing both the priority calculation method and GA as alternatives to the high-level policy, comparative results are displayed in Figs. 12 to 14. Notably, Fig. 13 indicates that the priority method takes approximately 33,000 steps to complete one construction episode. This result aligns closely with previous research [5], where agents required 32,800 simulation seconds to complete an episode. Such findings underscore the accuracy of the intelligent simulation facilitated by the low-level policy in replicating the temporal and spatial dynamics of construction.

Fig. 12 also reveals that the high-level policy, on average, requires about 30,000 steps to complete an episode. This indicates that the policy, trained through MAPPO algorithms in this CMDP, achieves approximately 10% greater overall construction efficiency compared to the alternatives. In contrast, while GA also aims to optimize construction duration, it tends to stabilize at around 35,000 steps and demonstrates larger oscillation amplitude. This variability stems from GA's nature as a pre-planned method, which may not always coincide with the agent's most optimal choices at any given moment, potentially leading to inefficient routing through congested areas or inopportune space occupancy.

Additionally, Fig. 13 compares the average number of steps taken by all seven agents to reach their targets in one episode during training. This comparison illustrates that the well-trained high-level strategy yields greater stability and a reduced total number of reaching steps, suggesting that agent movements under this strategy are less impacted by blocking effects. Fig. 14 presents a comparison of the average number of idle steps caused by spatial constraints for all seven agents in a single episode during training. The results demonstrate that the trained high-level policy is more effective in minimizing agent idleness due to space occupancy by other agents performing tasks, thereby indicating its superior capability in optimizing the temporal and spatial organization of tasking areas throughout the construction process.

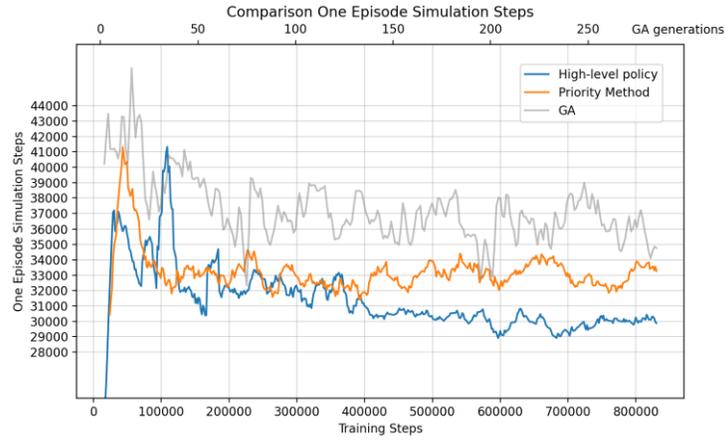

**Fig. 12.** Comparison of simulation steps in one episode with training steps grow

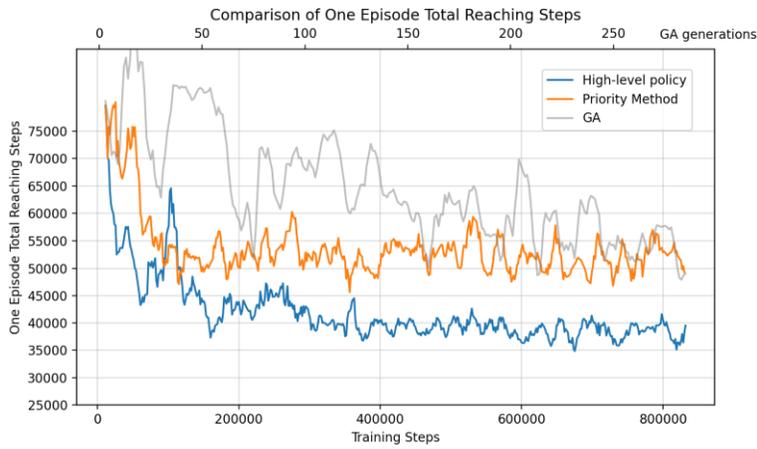

**Fig. 13.** Comparison of the seven agents' reaching steps in one episode

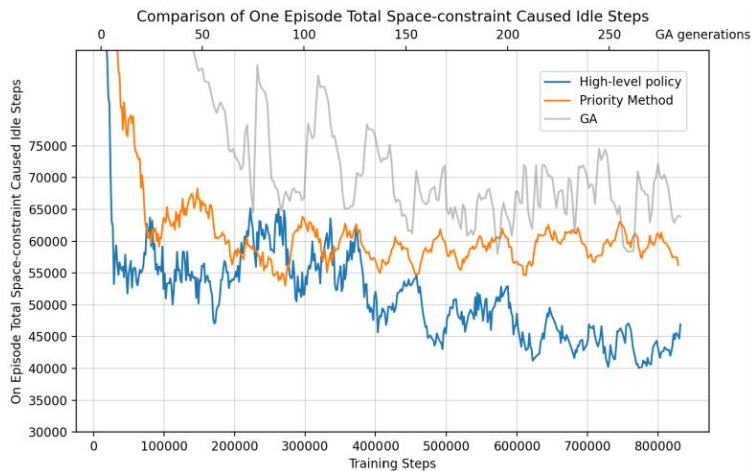

**Fig. 14.** Comparison of 7 agents' space-constrained caused idle steps in one episode

## 6. Discussion

Throughout its development and application in the case study, this CMDP has successfully accomplished a fine-grained construction simulation for agents, encompassing both mobility and action scales. It has also provided an effective training MDP focused on enhancing construction efficiency through the hierarchical strategy of crew agents. Distinguishing itself from existing research on Multi-Agent Reinforcement Learning (MARL) in construction [32][20][33], this CMDP goes beyond static site or path planning to transform the entire construction process into an MDP framework. Within this CMDP, agents are capable of observing a wide range of environmental dynamics and construction states, thereby enabling them to make decisions that are consistent with construction logic and aimed at achieving optimized construction goals.

The outcomes of the two-stage training process indicate that excluding task duration in the initial stage effectively fostered a highly rewarding target-reaching policy. This approach is akin to scenarios in existing multi-agent navigation studies [22][36], where agents are rewarded for reaching targets without delays caused by task execution. In the second stage, the lower-level strategy swiftly attained stable rewards, highlighting the high-level policy's capacity for continuous optimization of the spatiotemporal dynamics of task execution. This resulted in improved rewards and the establishment of a relatively stable strategy over time. These findings demonstrate that the integration of reinforcement learning into this CMDP has significantly accelerated knowledge discovery in the construction domain, thereby facilitating the development of innovative strategies to enhance construction efficiency.

Further comparisons with traditional task execution strategies and the GA method provide additional substantiation of the high-level policy's enhancements in construction efficiency. This is particularly evident in scenarios involving regional blockages and spatial constraints. These findings suggest that this CMDP, which derives agents' decision-making from a trained neural network, surpasses commonly employed heuristic methods in task planning [57]. Moreover, this framework distinguishes itself from related studies in multi-agent navigation, where agents' target selection is typically optimized based on factors such as distance [34][55] and collision condition status [56]. Instead, this CMDP effectively steers construction objectives by directing agents' target choices through observations encompassing both construction states and the effects of extended-duration construction tasks.

This methodology is specifically tailored for the floor construction process within a single work plane. To implement it effectively, the structural model, exclusively comprising floor components, must be imported. Additionally, all construction tasks and their interrelationships, constituting the construction environment, need to be meticulously defined and assigned to specific components. Furthermore, it is crucial to accurately model all required crew agents and equipment to establish a MAS simulation. Subsequently, this CMDP can be utilized to convert the MAS into a MDP, thereby incorporating RL to train agents for real-time decision-making in a dynamic construction environment. For more extensive construction projects, it is advisable to break them down into smaller subprocesses across different work planes. This CMDP can then be individually applied to each subprocess to simulate and optimize the respective construction processes.

However, it is important to acknowledge certain limitations of this CMDP. Its effectiveness is significantly dependent on the agents' ray observation outcomes. In complex scenarios characterized by overcrowding or the presence of blind spots among components, the method's performance might diminish. Although equipping agents with a "task map" or "memory storage" for the work plan could potentially mitigate this issue, such solutions fall outside the current research scope and may be explored in future studies. Additionally, this framework does not factor in unexpected events and uncertainties, such as delays and accidents, commonly encountered in construction. Future research is necessary to augment the agents' ability to adapt to and manage unforeseen circumstances effectively.

## 7. Conclusion

This paper aims to facilitate intelligent decision-making and simulation in construction scenarios by conceptualizing construction processes involving multiple agents with subjective decision-making as a series of dynamic target evaluation, reaching, and execution processes. It draws on insights from the fields of computer science, industry, and robotics to examine similar dynamic target selection and reaching methods. Through this review, key research challenges in developing MDP frameworks for construction scenarios are identified and addressed. These challenges include the integration of construction knowledge into agents' perception and decision-making processes, the development of policies capable of adapting to dynamically changing targets while fulfilling higher-level construction objectives, and the importance of considering the impact of sparse rewards during the training of agents within construction environments.

In response to these challenges, this paper presents the CMDP framework along with a novel two-stage training methodology. The CMDP framework's principal innovation resides in its capacity to enable agents to perceive their environment and make informed decisions grounded in construction knowledge, particularly focusing on various construction states. This framework streamlines observation modifications, thereby standardizing decision inputs and outputs across different crew agents. This standardization facilitates the use of a shared, centralized decision model for training purposes.

A crucial element of this framework is the implementation of a hierarchical policy model specifically designed for construction environments. The lower-level policy empowers agents to quickly reach high-value targets while adeptly navigating obstacles. Concurrently, the upper-level policy enhances various efficiency metrics within the overall construction process, unveiling innovative task execution patterns that improve both temporal and spatial resource allocation. Moreover, the introduction of a two-stage training approach plays a pivotal role in expediting the agents' proficiency in making target-reaching decisions within construction scenarios, thereby fostering quicker convergence and more effective decision-making.

The main contribution of this CMDP framework is its innovative portrayal of short-term construction processes as MDPs, thereby enabling the integration of reinforcement learning for the purpose of intelligent simulation and strategy development. This CMDP, implemented using the Unity platform, has undergone validation via a case study, demonstrating its effectiveness. The results confirm that the low-level policy successfully simulates the construction process intelligently, while comparative analysis with the priority calculation method indicates that the high-level policy substantially improves construction efficiency in various dimensions and facilitates the discovery of new construction strategies.

Furthermore, this framework holds significant promise for enhancing digital twins in construction settings. With the growing accessibility of detailed on-site data, made possible by advances in information technologies like computer vision, the intelligent simulation capabilities of this CMDP could be synergistically combined with real-time data in future studies. This integration aims to develop highly accurate virtual twins for construction activities, offering a transformative potential for the industry.

**Declaration of competing interest**

The authors declare that they have no known competing financial interests or personal relationships that could have appeared to influence the work reported in this paper.

**Acknowledgments**

The study is supported by the National Key R&D Program of China (Grant No. 2022YFC3801700) and the Top Discipline Plan of Shanghai Universities-Class I. A section of the original data about the construction task efficiency and duration is provided by the China Construction Eighth Engineering Division Corp. Ltd.